\documentclass[aps,prl,twocolumn,superscriptaddress]{revtex4}

\usepackage[latin9]{inputenc}
\usepackage{amsmath}
\usepackage{amssymb}
\usepackage{graphicx}
\usepackage{makecell}
\usepackage{array}
\usepackage{CJK}

\usepackage[unicode=true,pdfusetitle,
 bookmarks=false,
 breaklinks=false,pdfborder={0 0 1},backref=false,colorlinks=false]
 {hyperref}
\hypersetup{
 bookmarksnumbered=false,bookmarksopen=false}

\makeatletter
\@ifundefined{textcolor}{}
{%
 \definecolor{BLACK}{gray}{0}
 \definecolor{WHITE}{gray}{1}
 \definecolor{RED}{rgb}{1,0,0}
 \definecolor{GREEN}{rgb}{0,1,0}
 \definecolor{BLUE}{rgb}{0,0,1}
 \definecolor{CYAN}{cmyk}{1,0,0,0}
 \definecolor{MAGENTA}{cmyk}{0,1,0,0}
 \definecolor{YELLOW}{cmyk}{0,0,1,0}
}

\usepackage{color}

\@ifundefined{textcolor}{}{%
 \definecolor{BLACK}{gray}{0}
 \definecolor{WHITE}{gray}{1}
 \definecolor{RED}{rgb}{1,0,0}
 \definecolor{GREEN}{rgb}{0,1,0}
 \definecolor{BLUE}{rgb}{0,0,1}
 \definecolor{CYAN}{cmyk}{1,0,0,0}
 \definecolor{MAGENTA}{cmyk}{0,1,0,0}
 \definecolor{YELLOW}{cmyk}{0,0,1,0}
}

\usepackage{txfonts}

\makeatother

\begin{document}

\title{Observing the exponential growth of the eigenmodes in the absence of coalescence for a non-Hermitian circuit with an unavoidable inductor dissipation}

\author{Jiaming Zhang}
\affiliation{School of Science, Jiangxi University of Science and Technology, Ganzhou 341000, China}

\author{Wenxuan Song}
\affiliation{School of Science, Jiangxi University of Science and Technology, Ganzhou 341000, China}

\author{Hanhao Li}
\affiliation{School of Science, Jiangxi University of Science and Technology, Ganzhou 341000, China}

\author{Zhiye Kuang}
\affiliation{School of Science, Jiangxi University of Science and Technology, Ganzhou 341000, China}

\author{Zhihua Xiong}
\affiliation{School of Science, Jiangxi University of Science and Technology, Ganzhou 341000, China}

\author{Longwen Zhou}
\affiliation{College of Physics and Optoelectronic Engineering, Ocean University of China, Qingdao, China 266100}
\affiliation{Key Laboratory of Optics and Optoelectronics, Qingdao, China 266100}
\affiliation{Engineering Research Center of Advanced Marine Physical Instruments and Equipment of MOE, Qingdao, China 266100}

\author{Jie Liu}
\email[]{jliu@gscaep.ac.cn}
\affiliation{Graduate School of China Academy of Engineering Physics, Beijing 100193, China}
\affiliation{CAPT, HEDPS, and IFSA Collaborative Innovation Center of the Ministry of Education, Peking University, Beijing 100871, China}

\author{Wen-Lei Zhao}
\email[]{wlzhao@jxust.edu.cn}
\affiliation{School of Science, Jiangxi University of Science and Technology, Ganzhou 341000, China}

\begin{abstract}
We investigate, both experimentally and theoretically, the eigenmodes of an electronic circuit in which gain and loss $RLC$ resonators are coupled through a capacitor. Due to the unavoidable magnetic loss in the inductors, we find that the eigenmode coalescence no longer emerges in contrast to the conventional non-Hermitian systems with the spontaneous $\cal{PT}$-symmetry breaking. In particular, we find a transition from the exponential decay to exponential growth in the amplitude of the periodic voltage oscillations of the resonators. The transition occurs near the exceptional points of the non-Hermitian circuit without considering the dissipations in inductors. We introduce a small resistor of two orders of magnitude smaller than that of the $RLC$ resonators to mimic the energy dissipation in inductors and numerically solve the equivalent non-Hermitian Schr{\" o}dinger equation. The numerical results can well reproduce experimental observations. Our above findings unambiguously indicate that the exponential growth behavior beyond the exceptional points is robust against some unavoidable dissipative perturbations.
\end{abstract}
\date{\today}

\maketitle

\section{Introduction}
The fundamental problems arising from exceptional points (EPs) of non-Hermitian systems have received significant attention, both theoretically and experimentally, in different fields of physics, such as quantum metrology~\cite{XLYu24SA,Montenegro24arx}, optics~\cite{Miri19Sci,Ning25pra,Grelu12NP}, as well as condensed matter of physics~\cite{Bender98prl,Bender07RPP,Ashida20AP}. EPs act as branch points in complex plane, where encircling them in parameter space even induces a quantized accumulation of the Berry phase~\cite{Dembowski04pre,Dembowski01prl,Lee12pra}, uncovering the non-Hermitian topology~\cite{Liang13PRA,Tsubota22PRB,Bergholtz21PMP}. These topologically nontrivial singularities have profound implications for quantum transport in open systems~\cite{Ding22NRP,Longwen18prb,Huang24pra,Huichangg23pb}, leading, for instance, to the emergence of geometrically dependent skin effects in phononic crystals~\cite{QZhou23NC}, self-acceleration in dissipative quantum walks~\cite{PXue24NC}, and delocalization-localization transitions in quasicrystals~\cite{Lin22prl,Shan24prb110,Shan24prbl,Shan24prb109}, to name a few. EPs have been exploited to manipulate  the unidirectional reflectionless light propagation~\cite{YHuang17Nanoph} and perfect radiation in photonic devices~\cite{Inoue23PRApp}. Near the EPs, the square-root relationship between eigenenergy level splitting and system parameters amplifies significantly small perturbations, thereby dramatically enhancing the sensitivity of quantum sensors in various settings, including optical systems~\cite{Mao24SA},optomechanical sensors~\cite{Djorwe24PRR}, and electronic circuits~\cite{Chen24NSR}.

Recently, electronic circuits, offering precise control over key parameters, have emerged as powerful platforms for quantum simulations~\cite{Yang24PR,Jeon20PRL,Cheng22aps,Weixuan22nc,Weixuan21prl}. The high controllability of electronic components enables the effective emulation of diverse quantum systems and the exploration of exotic phenomena that are difficult to study in natural systems due to practical limitations. The coupled $RLC$ resonators, with balanced gain and loss, demonstrate the coalescence of eigenvalues of $\cal{PT}$-symmetric systems at EPs~\cite{Schindler24APL,Schindler12JPA}. By carefully designing $RLC$ resonator arrays with specific connectivity patterns, one can implement paradigm models in fundamental physics, including the quantum walk~\cite{Ezawa19PRB}, the Su-Schrieffer-Heeger lattice~\cite{Guo23apl}, and the Aubry-Andr{\' e} model~\cite{Halder23arx}. This paves the way for investigating novel physics like transient PT symmetry~\cite{Yang22prl}, topological insulating phases~\cite{ZWang23prl}, and non-Abelian Inverse Anderson transitions~\cite{WXZhang23prl}. Intrinsically, the versatility of electrical circuits, incorporating a tunable capacitor or inductor in the time domain, allows for the realization of Floquet $\cal{PT}$-symmetric systems~\cite{Chitsazi17PRL} and Floquet dissipative synthetic circuits~\cite{Montiel18CP}, opening new opportunities for engineering the non-equilibrium states of matter in non-Hermitian systems.

In typical circuits, inductors exhibit both unavoidable core losses and the Joule heating, leading to energy dissipation in an $RLC$ resonator. A fundamental issue is how such energy loss affects spontaneous $\cal{PT}$-symmetry breaking. Inspired by this, we investigate, both experimentally and theoretically, the effects of energy dissipation on the coalescence of the eigenmodes in coupled $RLC$ resonators, where energy loss in the inductors is modeled as an effective resistance in series with each inductor.
Without considering energy dissipation in the inductors, the system exhibits spontaneous $\mathcal{PT}$-symmetry breaking as a non-Hermitian parameter crosses the EPs~\cite{Schindler24APL,Schindler12JPA}. However, when energy dissipation is present, we find that the two real parts of eigenfrequency initially tend to merge as this non-Hermitian parameter increases but eventually develop a gap for sufficiently large values, indicating the breakdown of coalescence. For small values of this non-Hermitian parameter, the voltage of resonators exhibits periodic oscillation with an exponentially decaying amplitude over time. Interestingly, when this parameter exceeds the EPs of a $\cal{PT}$-symmetric circuit, the amplitude of the oscillation increases exponentially with time. Our above findings provide evidence of the robustness of exponential growth of eigenmodes against unavoidable dissipative perturbations, thereby offering significant implications for non-Hermitian physics.

The paper is organized as follows. In Sec.~\ref{MORes}, we describe the system and show the main results. A summary is presented in Sec.~\ref{Sum}.

\section{Model and main results}\label{MORes}

\begin{figure}[t]
\begin{center}
\includegraphics[width=8cm]{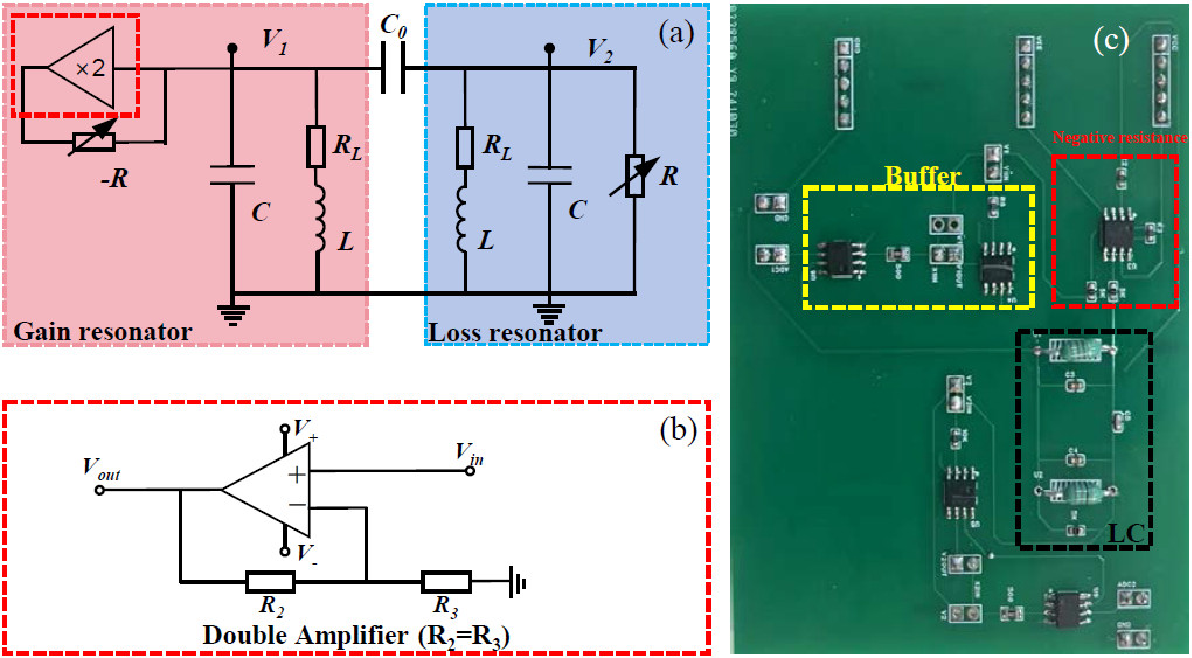}
\caption{(a) Schematic of $\cal{PT}$-symmetric $RLC$ resonator. The negative resistance $-R$ in the gain resonator is provided by feedback from a voltage-doubling buffer. The two resonators are coupled by the capacitor $C_0$. (b) Double amplifier with $R_2=R_3$. (c) Printed circuit board of the $\cal{PT}$-symmetric $RLC$ resonator. The buffer consists of high-impedance LF356 voltage followers, designed to isolate the influence of measurement instruments on the $RLC$ resonator.
\label{SchDiagr}}
\end{center}
\end{figure}

The experimental setup consists of two parallel $RLC$ resonators, coupled by a capacitor $C_0$, and integrated on a printed circuit board (PCB) [see Figs.~\ref{SchDiagr}(a) and (c)]. The loss resonator, highlighted in the blue region, has a positive resistance $R$, while the gain resonator, highlighted in the red region, features a negative resistance $-R$. The negative resistance is implemented using a voltage-doubling buffer [see Fig.~\ref{SchDiagr}(b)], where the output current follows the relation $I_{\text{out}} = (V - 2V)/{R} =- {V}/{R}$ effectively realizing a negative resistance. In our experiment, we use an axial fixed inductor with inductance $L = 773 \, \mu\text{H}$, SMD capacitors with capacitances $C = 4.78 \, \text{nF}$ and $C_0 = 1.6 \, \text{nF}$, and tunable resistors $R$ ranging from 800 $\Omega$ to 15 $\text{k}\Omega$. The difference in resistance between the positive and negative resistors is less than $0.5\%$. The unavoidable magnetic loss in the inductor, accompanied by energy dissipation, disrupts the exact balance between energy gain and loss, causing a deviation from the ideal $\cal{PT}$-symmetric system. To compensate for this energy dissipation, a gain component is introduced in each coil to realize a
$\cal{PT}$-symmetric circuit~\cite{Schindler24APL,Chitsazi17PRL,Schindler12JPA}.
In this work, we introduce a resistance $R_L$ in the coil to model the energy dissipation of each inductor.

Based on Kirchhoff's laws, one can obtain the equations for the current $I$ and voltage $V$ of gain and loss nodes
  \begin{equation}\label{KircffLaw}
  	\begin{aligned}
  		&I_n^C+I_n^R+I_n^L+I_n^{C_0}=0\;,&\\
  		&V_n=R_LI_n^L+L\frac{dI_n^L}{dt}\;,&
  	\end{aligned}  	
  \end{equation}
where the superindices $R$, $L$, $C$, and $C_0$ indicate quantities   associated with the resistor, inductor, capacitor and coupled capacitor, respectively. The subscripts denote the gain ($n=1$) and loss ($n=2$) resonators. Here, the current of the resistors and capacitors take the form $I_n^R=(-1)^nV_n/{R}$, $I_1^{C_0}=C_0({dV_1}/{dt}-{dV_2}/{dt})$, $I_2^{C_0}=-I_1^{C_0}$, and $I_n^C=C{dV_n}/{dt}$. For brevity, we introduce a set of dimensionless quantities $c={C_0}/{C}$, $k={R_L}/{R}$, $\gamma = {R}^{-1}\sqrt{{L}/{C}}$, and $\gamma_l=R_L\sqrt{{C}/{L}}$. Dimensionless time is naturally defined as $\tau=\omega_0 t$, where $\omega_0=1/\sqrt{LC}$ represents the resonant frequency. Straightforward calculations yield the relations
 \begin{equation}\label{VEquation}
 	\begin{aligned}
 	\ddot{V}_1=&\dot{V}_1\frac{\gamma-(1+c)\gamma_l}{c+1}+ V_1\frac{k-1}{c+1}+\ddot{V}_2\frac{c}{c+1}+\dot{V}_2\frac{c\gamma_l}{c+1}\;,\\
 	\ddot{V}_2=&-\dot{V}_2\frac{\gamma+(1+c)\gamma_l}{c+1}- V_2\frac{k+1}{c+1}+\ddot{V}_1\frac{c}{c+1}+\dot{V}_1\frac{c\gamma_l}{c+1}\;.
 	\end{aligned}
 \end{equation}
where $\dot{V}_i=dV_i/d\tau$ and $\ddot{V}_i=d^2V_i/d\tau^2$ with $i=1,2$.
By introducing the vector $\psi = (V_1,V_2,\dot{V}_1,\dot{V}_2)$, we can rewrite above equations in analogy to the Schr{\" o}dinger equation (with $\hbar=1$)
 \begin{equation}\label{SchEquation}
	i\frac{d\psi}{d\tau}=\text{H}\psi\;,
 \end{equation}
with the Hamiltonian
\begin{equation}\label{HamilMatrix}
 	\begin{aligned}
 			&\text{H}=i	\begin{pmatrix}
 				    0 & \mathbb{I} \\
 				    M_a & M_b
 					\end{pmatrix}\;,\\
 	\end{aligned}
 \end{equation}
where $0$ is the zero matrix, $\mathbb{I}$ is the identity matrix, and matrices $M_a$ and $M_b$ are given by
\begin{equation}\label{AMatrix}
 	\begin{aligned}
 		&M_a=\frac{1}{1+2c}\begin{bmatrix}
 			{(1+c)(k-1)} & -c(k+1)  \\
 			{c(k-1)} & -(1+c)(k+1)
 		\end{bmatrix}\;,\\
 	\end{aligned}
 \end{equation}
and
\begin{equation}\label{AMatrix}
 	\begin{aligned}
 		&M_b=\frac{1}{1+2c}\begin{bmatrix}
			 	 (1+c)\gamma-(1+2c)\gamma_l & -c\gamma \\
			 	 c\gamma & -(1+c)\gamma-(1+2c)\gamma_l
 		   \end{bmatrix}\;.
 	\end{aligned}
 \end{equation}

\begin{figure}[t]
\begin{center}
\includegraphics[width=8cm]{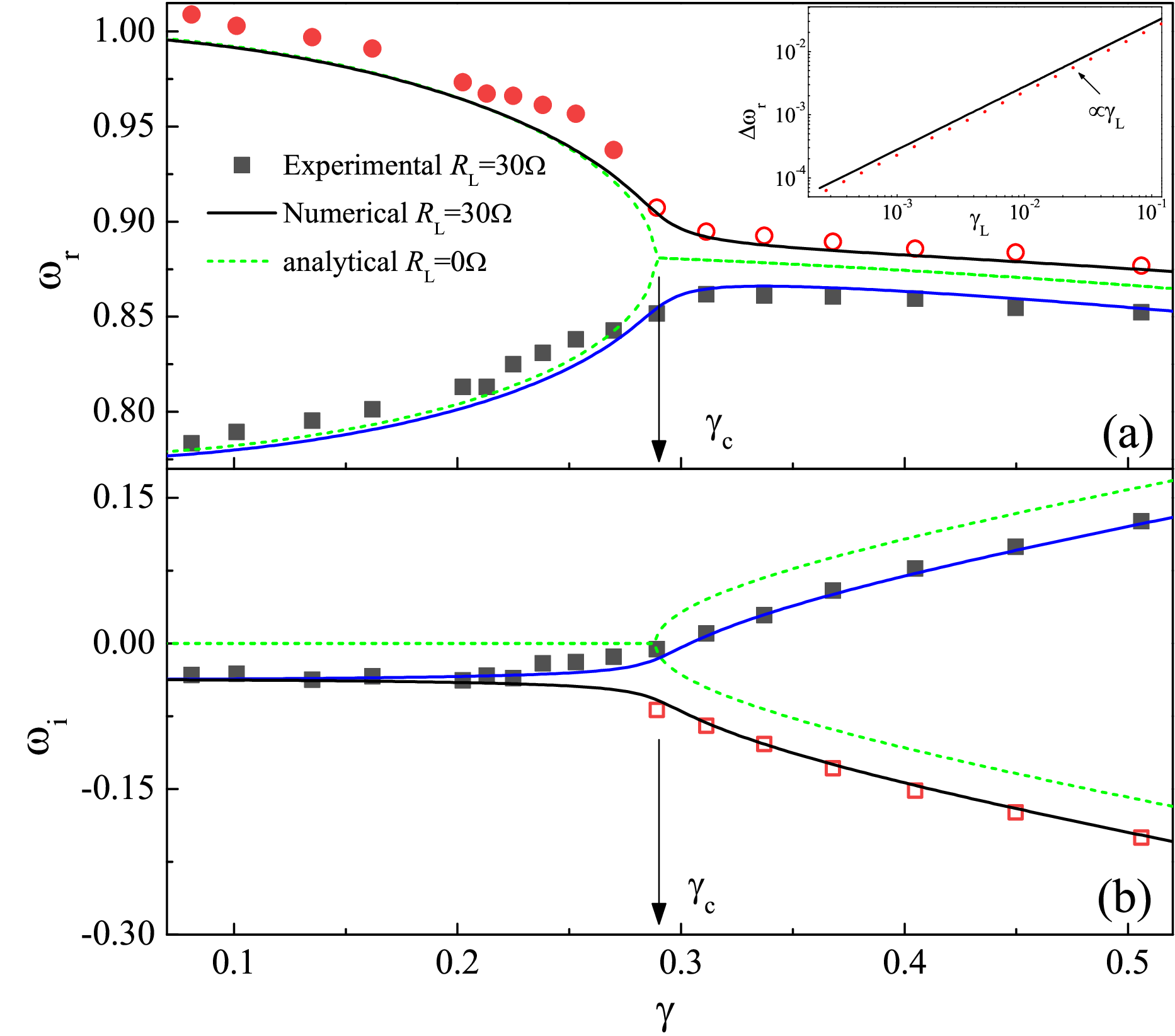}
\caption{Experimentally measured eigenfrequencies $\omega=\omega_r+i\omega_i$ versus $\gamma$. The real $\omega_r$ and imaginary parts $\omega_i$ are shown in (a) and (b), respectively. Solid lines indicate numerical results of $\omega$ for $\gamma_L =0.0738$. Green-dashed lines are analytical prediction of $\omega$ for $\gamma_L=0$, manifesting an exponential point at $\gamma_c=0.29097$. Open circles in (a) are reflections of the experimental data (lower branch) with respect to $\omega_r\approx 0.864$. Inset in (a): The gap of the real part $\Delta \omega_r$ versus $\gamma_L$ for $\gamma > \gamma_c$. Red-dashed line indicates the linear fitting, i.e., $\Delta \omega_r\approx 0.23\gamma_L$.
Open squares in (b) are reflections of the experimental data (upper branch) with respect to $\omega_i=-0.03706$.
\label{EEnergy}}
\end{center}
\end{figure}

For $\gamma_l=0$, we can obtain analytically the eigenvalue of the Hamiltonian
\begin{equation}
			\omega_{1,3} = \sqrt{\frac{2(1+c)-\gamma^2\pm\sqrt{[\gamma^2-2(1+c)]^2-4(2c+1)}}{2(2c+1)}}\;,
		\end{equation}
and $\omega_{2,4} = -\omega_{1,3} $. The two positive eigenvalues $\omega_1$ and $\omega_3$ coalesce at the exceptional point $\gamma_c = \sqrt{2(1+c)+\sqrt{1+2c}}$, beyond which the eigenvalues acquire imaginary components, indicating the breaking of $\cal{PT}$-symmetry (see Fig.~\ref{EEnergy}).

In realistic scenarios, nonzero $\gamma_l$ arises from the intrinsic energy dissipation of the inductor $L$. This renders analytical derivation of the eigenvalues for the non-Hermitian Hamiltonian $\rm{H}$ intractable. We therefore perform numerical diagonalization of $\rm{H}$ to resolve its complex eigenvalue $\omega=\omega_r+ i\omega_i$. Our results show that for very small $\gamma_l$ (i.e., $\gamma_l=0.0738$ in Fig.~\ref{EEnergy}), the real eigenvalues $\omega_r$ have very slight difference from its $\gamma_l = 0$ counterpart when $\gamma<\gamma_c$. Intrinsically, however, the two $\omega_r$ branches remain non-degenerate at $\gamma = \gamma_c$ and develop a finite gap $\Delta\omega_r$ for $\gamma > \gamma_c$, signaling the breakdown of spectral coalescence in the $\mathcal{PT}$-symmetric systems. We find that the gap linearly increases with $\gamma_L$ [see the inset in Fig.~\ref{EEnergy}(a)]. Due to the unavoidable energy dissipation of the coil $L$, the eigenvalues are complex, i.e., $\omega_{j}=\omega_{r,j}+i\omega_{i,j}$ for $j=1,2$ [see Fig.~\ref{EEnergy}(b)]. The imaginary parts of the eigenvalues remain nearly constant and show negligible deviation from each other when $\gamma < \gamma_c$, but they diverge noticeably as $\gamma > \gamma_c$.

We further perform experimental measurements of the eigenfrequencies of the coupled $RLC$ resonators. A comparison between $\omega$ and the numerical results is shown in Fig.~\ref{EEnergy}. One can see that for $\gamma < \gamma_c$, the experimentally measured values of both $\omega_r$ and $\omega_i$ agree well with numerical simulations based on the model in Eq.~\eqref{HamilMatrix}. Note that we can extract only one imaginary component of the eigenfrequency because the difference between the two $ \omega_{i}$ values is too small. For $\gamma > \gamma_c$, the eigenmode of the electronic circuit associated with a negative imaginary part $\omega_i$ decays exponentially. Therefore, only the eigenfrequency with $\omega_i>0$ is experimentally accessible, and it is in perfect agreement with numerical results.
\begin{figure}[t]
\begin{center}
\includegraphics[width=8cm]{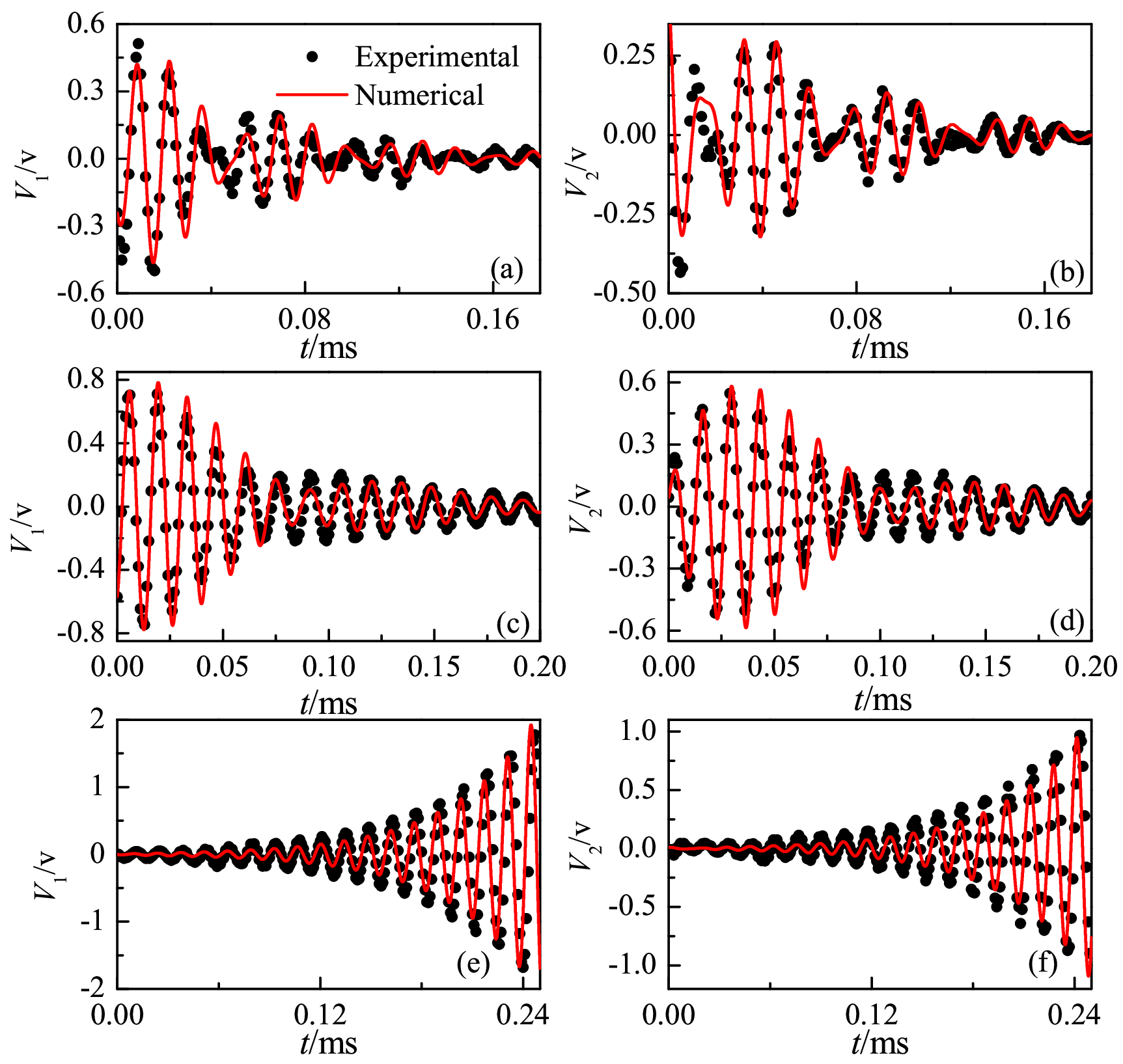}
\caption{(a) Time dependence of the both $V_1$ and $V_2$ for $\gamma=0.0815$ (top panels) 0.2545 (middle panels), and 0.3133 (bottom panels). Solid lines indicate our numerical results.\label{Voltage}}
\end{center}
\end{figure}

We determine the eigenfrequency by applying a Fourier transform to the time evolution of the voltage $V_j$ at each node. Figure~\ref{Voltage} shows the experimentally measured values of $V_j$ for different $\gamma$. One can see that for $\gamma<\gamma_c$, the $V_j$ periodically oscillates with time. The envelope of this oscillation is also periodic, while exhibiting an exponentially decaying amplitude [see Figs.~\ref{Voltage}(a) and (b)]. This reveals the existence of two real parts in the eigenfrequency. The rate of the exponential decay of the envelop's amplitude corresponds to the imaginary parts of the eigenfrequency. For $\gamma\approx \gamma_c$, the amplitude of the envelope of $V_j$ is significantly larger compared to smaller values of $\gamma$. Interestingly, for $\gamma>\gamma_c$, the time evolution of $V_j$ exhibits periodic oscillations with a single frequency, and the amplitude increases exponentially with time. In this case, the oscillation frequency and the growth rate of the amplitude are $\omega_r$ and $\omega_i$, respectively. For comparison, we also employ the fourth-order Runge-Kutta algorithm to numerically solve Eq.~\eqref{SchEquation} and obtain the time evolution of $V_j$. Our numerical results can well reproduce the experimentally measured values of $V_j$, confirming the validity of our model. This consistency is clear evidence that our introduction of $R_L$ to quantify energy dissipation in the coil $L$ is physically reasonable.

Non-Hermiticity is now widely recognized as an important extension of conventional quantum mechanics, providing a framework for describing open quantum systems. Notable phenomena, including the non-Hermitian many-body problem~\cite{Xie19foP}, biorthogonal dynamical phase transitions~\cite{Jing24prl}, and non-reciprocal adiabatic tunneling~\cite{Chang25cpb}, have demonstrated the fundamental significance of non-Hermitian physics. To implement $\mathcal{PT}$-symmetric circuits, gain components should be utilized to compensate for energy loss in inductors~\cite{Schindler24APL,Schindler12JPA}. In this situation, for $\gamma < \gamma_c$, the voltage of the active $RLC$ resonators exhibits periodic oscillations due to energy gain. However, the unavoidable magnetic loss in inductors, which introduces small energy dissipation, leads to the exponential decay of voltage oscillations, corresponding to a negative value of $\omega_i$. Intrinsically, the eigenmodes of the voltage signal in our system exhibit a transition to exponential growth with time as the non-Hermitian parameter $\gamma$ crosses the EPs, indicating the robustness of exponential energy gain against perturbative dissipation.

\section{Conclusion and discussions}\label{Sum}

The non-Hermiticity of electronic circuits generally arises from dissipative elements that introduce energy loss or gain, breaking energy conservation. In this work, we experimentally realize an electronic circuit where gain and loss $RLC$ resonators are coupled by a capacitor. The gain side is implemented using a voltage-doubling buffer, which provides feedback current with the relation $I_{out} = -V/R$, effectively achieving a negative resistance. Energy dissipation in the inductors is present in both the gain and loss $RLC$ resonators. Theoretically, we model this energy loss by introducing resistors $R_L$ in each inductor, thereby quantifying energy dissipation precisely. For $R_L=0$, the spontaneous $\cal{PT}$-symmetry breaking is controlled by a dimensionless parameter $\gamma$, with the EP occurring at $\gamma=\gamma_c$. We find that $V_j$ exhibits periodic oscillations over time, with its amplitude decaying exponentially for $\gamma < \gamma_c$ and increasing exponentially for $\gamma > \gamma_c$. Correspondingly, the imaginary part $\omega_i$ of the eigenfrequencies is negative for $\gamma < \gamma_c$ and positive for $\gamma > \gamma_c$. In addition, the two branches of $\omega_i$ diverge as $\gamma$ increases. The two real parts of the eigenfrequency converge as $\gamma$ increases but retain a gap for $\gamma > \gamma_c$. Therefore, the coalescence of eigenmodes is broken by the energy dissipation of inductors. Our finding that $\omega_r$ exhibits a gap for $\gamma > \gamma_c$  may represent a unique feature of non-Hermitian circuits, whose implications on sensing and wave manipulation warrant further investigation.

\section*{ACKNOWLEDGMENTS}
This work is supported by the National Natural Science Foundation of China (Grant No. 12365002, 12065009 and 12364013), the Natural Science Foundation of Jiangxi province (Grant No. 20224ACB201006 and 20224BAB201023). Jie Liu is supported by the NSAF (Contract No. U2330401).

\end{document}